\documentclass[10pt]{article}

\usepackage[dvips]{graphicx,color}
\usepackage[psamsfonts]{amssymb}
\usepackage{latexsym}
\usepackage{amsmath}
\usepackage{amsthm}
\usepackage{amsfonts}
\usepackage[T1]{fontenc}
\usepackage{textcomp}

\begin{document}
\begin{center}
{\Large Improving the efficiency of decoding quantum error correction code}
\end{center}
\begin{center}
Kenichiro Furuta\\
kenichiro.furuta@toshiba.co.jp\\   
Corporate Research \& Development Center, Toshiba Corporation\\
1, Komukai-Toshiba-cho, Saiwai-ku, Kawasaki, 212-8582, JAPAN\\
\end{center}
{\bf abstract} \hspace{0.5cm}
To improve the efficiency of the encoding and the decoding is the important problem in the quantum error correction.
In \cite{explofgott}, a general algorithm for decoding the stabilizer code\cite{firststabilizer} is shown.
This paper will show an decoding which is more efficient for some codes.
The proposed decoding as well as the conventional decoding consists of the eigenvalue output step and
the entanglement dissolution step.

The proposed decoding outputs a part of the eigenvalues into a part of the code qubits
in contrast to the conventional method's outputting into the ancilla.
Besides, the proposed decoding dissolves a part of the entanglement in the eigenvalue output step
in contrast to the conventional method which does not dissolve in the eigenvalue output step.
With these improvements, the number of gates was reduced for some codes.\\
{\bf keywords} \hspace{0.5cm} quantum error correction, stabilizer code, quantum computer
\vspace{0.5cm}
\section{Introduction}
It is important to controll the quantum state in order to realize 
the quantum computation and the quantum cryptography.
The quantum error correction\cite{firststabilizer} is proposed for controlling the quantum state.

In Chap.4 of \cite{explofgott}, a general algorithm of decoding the stabilizer code
is shown and the efficiency of the algorithm is also discussed.
This paper shows more efficient decoding algorithm by specifying some codes.

Paper\cite{correctanddecode} proposed that the simultaneous correction and decoding
leads to the decrease of the number of the gates.
However, for the quantum code word whose error is estimated to be small, 
it is sufficient to perform only the decoding.
For example, the encoded state in the quantum memory will be corrected 
every constant time interval.
When the quantum state is taken out from the quantum memory soon after the correction,
we have only to decode. The algorithm that performs only the decoding is valid in cases as this.
This is why this paper focuses the efficient decoding circuit.
\section{Summary of the quantum error correction}
Almost all quantum error correction codes are constructed as the clifford code. 
The stabilizer code\cite{firststabilizer} is the subset of the clifford code\cite{clifford}
and the CSS code is the subset of the stabilizer code.

In this paper, only the stabilizer codes whose code words are expressed on qubits are dealt with.
The possible unitary operator can be written as the superposition of the error basis.
The error basis forms the group which is called the error group.
The Pauli operators are as
$I=\left[ \begin{array}{ccc}1&0\\0&1
\end{array} \right],
\sigma_{x}=\left[ \begin{array}{ccc}0&1\\1&0
\end{array} \right],
\sigma_{z}=\left[ \begin{array}{ccc}1&0\\0&-1
\end{array} \right],
\sigma_{x}\sigma_{z}.$
With the Pauli operators, the error group is written as
$E^{n}= \{ \pm N_{1} \otimes \cdots \otimes N_{n} \mid
N_{1},\cdots,N_{n} \in \{ I,\sigma_{x},\sigma_{z},\sigma_{x}\sigma_{z} \} \}.$
Let $P$ be an abelian subgroup of the error group. 
Let $Q$ be an eigenspace of $P$.
The eigenspace is a set of eigenstates whose eigenvalues are equivalent each other for arbitrary operators in $P$.
The code word is an element of $Q$.
An abelian subgroup $P$ can be represented with some generators as 
$P=<M_{1},M_{2},\cdots,M_{f}>$.

With the code word whose minimum distance is $d$,
error basis whose weight is no more than $t=\lfloor(d-1)/2\rfloor$ can  be corrected. 
The detail of the definition of the minimum distance is 
written in \cite{firststabilizer}\cite{explofgott}.

The code word is written as follows.
Let $\alpha_{0}|0\rangle+\alpha_{1}|1\rangle+\cdots+\alpha_{g-1}|g-1\rangle$
be a quantum state before encoding, where
$g$ is a natural number, and $\alpha_{j}$ is a complex number.
Then, the code word is expressed as
$|\Psi\rangle=\alpha_{0}|\Psi_{0}\rangle+\alpha_{1}|\Psi_{1}\rangle+\cdots
+\alpha_{g-1}|\Psi_{g-1}\rangle,$
where $|\Psi_{j}\rangle= \sum_{k}\beta_{j,k}|\Psi_{j,k}\rangle$,
and $|\Psi_{j,k}\rangle=|\Psi_{j,k}\rangle_{1} \otimes \cdots \otimes |\Psi_{j,k}\rangle_{n}$,
and $\beta_{j,k}$ is a complex number.
Let qubits which express the code word at the beginning of the decoding be called the code qubits.

The code word of the 5 qubit code is written as follows.
Let $|0\rangle$ be encoded into
$|\Psi_{0}\rangle=\frac{1}{\sqrt{2}}(|00000\rangle+|10010\rangle+|01001\rangle+|10100\rangle
+|01010\rangle+|00101\rangle-|11110\rangle-|01111\rangle
-|10111\rangle-|11011\rangle-|11101\rangle-|01100\rangle
-|00110\rangle-|00011\rangle-|10001\rangle-|11000\rangle)$,
and $|1\rangle$ be encoded into $|\psi_{1}\rangle=(\sigma_{x} \otimes \sigma_{x} \otimes \sigma_{x} \otimes \sigma_{x} \otimes \sigma_{x})|\psi_{0}\rangle$.

The abelian subgroup, $P$, is called the stabilizer and is represented by some generators.
The eigenvalues for these generators index the eigenspace.
For example, the generators for the 5 qubit code are shown as
$M_{1}=\sigma_{x} \otimes \sigma_{z}  \otimes \sigma_{x}  \otimes \sigma_{z}  \otimes I$,
$M_{2}=I  \otimes \sigma_{x} \otimes \sigma_{z} \otimes \sigma_{z} \otimes \sigma_{x}$,
$M_{3}=\sigma_{x} \otimes I \otimes \sigma_{x} \otimes \sigma_{z} \otimes \sigma_{z}$,
$M_{4}=\sigma_{z} \otimes \sigma_{x} \otimes I \otimes \sigma_{x} \otimes \sigma_{z}$,
and other code parameters are shown as
$\bar{X}=\sigma_{x} \otimes \sigma_{x} \otimes \sigma_{x} \otimes \sigma_{x} \otimes \sigma_{x}$,
$\bar{Z}=\sigma_{z} \otimes \sigma_{z} \otimes \sigma_{z} \otimes \sigma_{z} \otimes \sigma_{z}$.

The 5 qubit code is written as follows with the stabilizer.
$|\psi_{0}\rangle$ is written as $\sum_{M \in P}M|00000\rangle
=|00000\rangle+M_{1}|00000\rangle+\cdots+M_{4}|00000\rangle
+M_{1}M_{2}|00000\rangle+\cdots+M_{3}M_{4}|00000\rangle
+M_{1}M_{2}M_{3}|00000\rangle+\cdots+M_{2}M_{3}M_{4}|00000\rangle+M_{1}M_{2}M_{3}M_{4}|00000\rangle$ 
and $|\psi_{1}\rangle$ is written as $|\psi_{1}\rangle=\bar{X}|\psi_{0}\rangle$.
The code word also satisfies $\bar{Z}|\psi_{0}\rangle=|\psi_{0}\rangle,\bar{Z}|\psi_{1}\rangle=-|\psi_{1}\rangle$.

I describe the conventional algorithm for the encoding and the decoding in the following.
The general algorithm of the encoding is written in \cite{explofgott}.
We will illustrate the summary of this method. 

The code word are specified by some parameters, $M_{i},\bar{X},\bar{Z}$, as shown above.
The generators $M_{i}$ and $\bar{X}$ are transformed into the standard form.

The algorithm for the encoding is shown in Chap.4 of \cite{explofgott}
with the matrix that corresponds to $\bar{X}$ and $M_{i}$.
The outline of the algorithm for the encoding is written as follows.
See the Chap.4 of \cite{explofgott} in detail.\\
1.\hspace{0.2cm}The WH gate is applied on a part of qubits.\\
2.\hspace{0.2cm}The operations corresponding to the standard form of $\bar{X}$ are applied.\\
3.\hspace{0.2cm}For each $i$, the operations corresponding to the standard form of $M_{i}$ are applied.

Let $U_{enc}$ be the unitary operator for the encoding.
Then, the unitary operator for decoding can be $U_{enc}^{-1}$,
which is the inverse operator of $U_{enc}$.
$U_{enc}$ is needed to encode qubits.
However, we need not apply $U_{enc}^{-1}$ to decode.
We can decode with the less number of gates\cite{explofgott}.
A method for improving the efficiency of the decoding is shown in \cite{explofgott}.
I will illustrate the detail of this algorithm.
Let $\sum_{i}\alpha_{i}|i\rangle$ be encoded 
into $\sum_{i}\alpha_{i}|\Psi_{i}\rangle$.
Then, $|\Psi_{i}\rangle$ does not depend on $\alpha_{i}$.
By applying $U_{enc}^{-1}$, the code changes into 
$(\sum_{i}\alpha_{i}|i\rangle)\otimes|0\rangle\otimes\cdots\otimes|0\rangle$.
Without applying $U_{enc}^{-1}$, the code word can be changed into
$|\Psi_{0}\rangle \otimes (\sum_{i}\alpha_{i}|i\rangle)$ by the following method.
This method consists of 2 steps. The 1st step is called the eigenvalue output step
and the 2nd step is called the entanglement dissolution step.

In the eigenvalue output step, the eigenvalue that indicates 
which $|\Psi_{i}\rangle$ each basis is contained in is output.
These values are the eigenvalues of $\bar{X},\bar{Z}$.
The code qubits are transformed into $\sum_{i}\alpha_{i}|\Psi_{i}\rangle\otimes|i\rangle$.

In the entanglement dissolution step, gates for dissolving the entanglement between
the qubits which represent $\sum_{i}\alpha_{i}|i\rangle$ and other qubits are applied.
Then, the code qubits are transformed into
$\sum_{i}\alpha_{i}|\Psi_{0}\rangle\otimes|i\rangle
=|\Psi_{0}\rangle\otimes( \sum_{i}\alpha_{i}|i\rangle)$.

\begin{figure}[ht]
\includegraphics[width=12cm,clip]{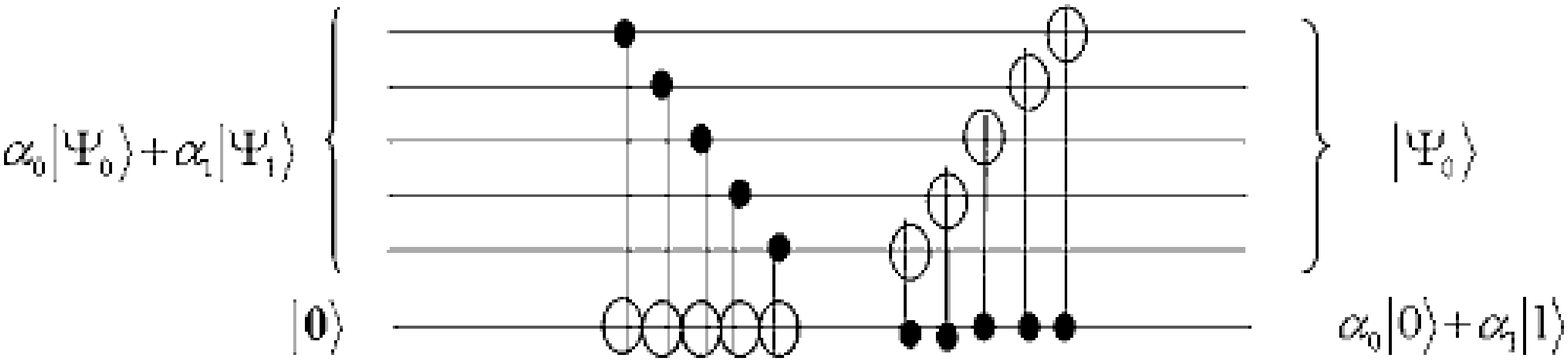}
\caption{A conventional method for the 5 qubit code}
\label{fig:5qbsofar}
\end{figure}

In the decoding method above, the resulting state of the decoding is 
$|\Psi_{0}\rangle\otimes(\sum_{i}\alpha_{i}|i\rangle)$
and $|\Psi_{0}\rangle$ is left as it was in the initial stage.
Not applying the operation of transforming $|\Psi_{0}\rangle$ 
into $|0\cdots0\rangle$ leads to more efficient decoding.
So, it seems that leaving $|\Psi_{0}\rangle$ is the best choice.
An example for the 5 qubit code is shown in Fig.\ref{fig:5qbsofar}.
\section{A proposition for improving the efficiency of the decoding}
\subsection{Change in the eigenvalue output location}
The proposition in this paper is the algorithm for improving the efficiency of the conventional deoding algorithm.
We show here that leaving $|\Psi_{0}\rangle$ is not the best choice and changing a part of qubits 
expressing $|\Psi_{0}\rangle$ is better choice for some codes.

The conventional algorithm output the eigenvalues on the ancilla.
The proposed algorithm outputs a part of the eigenvalues on a part of the code qubits.
Output of the eigenvalue into the code qubits leads to the decrease of the
number of times of moving information and the number of the gates.

\begin{figure}[ht]
\includegraphics[width=12cm,clip]{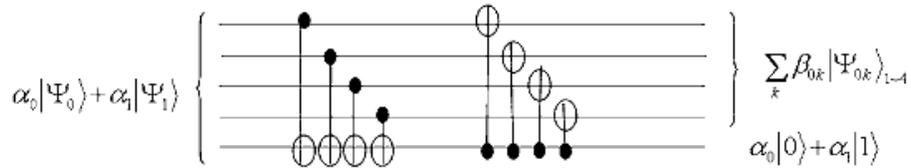}
\caption{A proposed method for the 5 qubit code}
\label{fig:5qbpropose}
\end{figure}

An example of the circuit for the 5 qubit code which is shown in Fig.\ref{fig:5qbpropose}
outputs the eigenvalue on a part of the code qubits.
\subsection{The dissolution of the entanglement}
The less number of qubits are entangled after the eigenvalue output step,
the less number of gates are applied in the entanglement dissolution step.
So, it may be useful to attempt to decrease the number of qubits that are entangled
at the end of the eigenvalue output step.

The dissolution of the entanglement is defined in the following.
A quantum state, $\alpha_{0}|0\rangle+\alpha_{1}|1\rangle$,
is encoded into $\alpha_{0}|\Psi_{0}\rangle+\alpha_{1}|\Psi_{1}\rangle$, 
where $|\Psi_{j}\rangle=\sum_{k} \beta_{k}|\Psi_{jk}\rangle$
and $\beta_{k}=\beta_{0k}=\beta_{1k}$.
The code words that are dealt with here satisfy the following relationship 
for $|\Psi_{0k}\rangle$ and $|\Psi_{1k}\rangle$.
Applying NOT gate on all code qubits, which is called the reversing, 
translate $|\Psi_{0k}\rangle$ into $|\Psi_{1k}\rangle$.
Thus, $|\Psi_{0k}\rangle$ and $|\Psi_{1k}\rangle$ are symmetric with respect to reversing.
Then, $|\Psi_{0k}\rangle$ and $|\Psi_{1k}\rangle$ are called the symmetric basis pair.
The dissolution of the entanglement for $1 \sim l-1$th code qubits in the symmetric basis pair means that
\begin{eqnarray}
\sum_{k} \beta_{k}
( \alpha_{0}|\Psi_{0k}\rangle_{1 \sim l-1}|\Psi_{0k}\rangle_{l \sim m-1}|0\rangle_{m}
+\alpha_{1}|\Psi_{0k}\rangle_{1 \sim l-1}|\Psi_{1k}\rangle_{l \sim m-1}|1\rangle_{m}) \notag \\
=\sum_{k} \beta_{k}|\Psi_{0k}\rangle_{1 \sim l-1} \otimes
\left( \alpha_{0}|\Psi_{0k}\rangle_{l \sim m-1}|0\rangle_{m}
+\alpha_{1}|\Psi_{1k}\rangle_{l \sim m-1}|1\rangle_{m} \right). \notag
\end{eqnarray}

In order to dissolve the entanglement in the eigenvalue output step,
I use the linked CNOT chain defined as follows.
A linked CNOT chain is a sequence of multiple CNOT gates,
in which the target qubit of a CNOT gate is equal to the controll qubit of the next CNOT gate. 
Besides, for each CNOT gate, its target qubit
is different from the target qubit of other CNOT gates
and from the controll qubit of the 1st CNOT gate of the linked CNOT chain.

The entanglement is dissolved at the termination of the eigenvalue output step 
for some code qubits in the symmetric basis pair
by employing the linked CNOT chain in the eigenvalue output step.

Let $|a_{1},a_{2},a_{3},a_{4},a_{5}\rangle$ and $|b_{1},b_{2},b_{3},b_{4},b_{5}\rangle$ be 2 basis that satisfy
$b_{i}=\bar{a_{i}}$ for $i=1,...,5$, where $a_{i},b_{i} \in \{0,1\}$ 
and $\bar{a_{i}}=0(\bar{a_{i}}=1)$ for $a_{i}=1(a_{i}=0)$.
Let $|a_{1}',a_{2}',a_{3}',a_{4}',a_{5}'\rangle(|b_{1}',b_{2}',b_{3}',b_{4}',b_{5}'\rangle)$
be the state of $|a_{1},a_{2},a_{3},a_{4},a_{5}\rangle(|b_{1},b_{2},b_{3},b_{4},b_{5}\rangle)$
after the eigenvalue output step.

\begin{figure}[ht]
\includegraphics[width=12cm,clip]{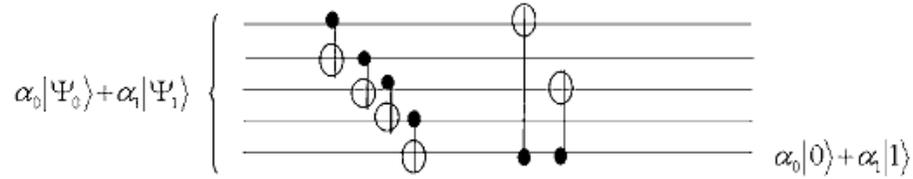}
\caption{A proposed method for the 5 qubit code which is more efficient than Fig.\ref{fig:5qbpropose}}
\label{fig:5qbleast}
\end{figure}

We illustrate the change of the state of these 2 basis in Fig.\ref{fig:5qbleast}.
After having performed first 4 gates,
$a_{1}'=a_{1},a_{2}'=a_{1}+a_{2},a_{3}'=a_{1}+a_{2}+a_{3},
a_{4}'=a_{1}+a_{2}+a_{3}+a_{4},a_{5}'=a_{1}+a_{2}+a_{3}+a_{4}+a_{5}$
and 
$b_{1}'=b_{1},b_{2}'=b_{1}+b_{2},b_{3}'=b_{1}+b_{2}+b_{3},
b_{4}'=b_{1}+b_{2}+b_{3}+b_{4},b_{5}'=b_{1}+b_{2}+b_{3}+b_{4}+b_{5}.$
By comparing $a_{i}'$ with $b_{i}'$, we can know
$b_{i}'=\bar{a_{i}'}$ for $i=1,3,5$ and 
$b_{i}'=a_{i}'$ for $i=2,4$.
The additions above are performed in GF(2).

After applying the linked CNOT chain on the symmetric basis pair,
values are equivalent each other in the symmetric basis pair
for the controll qubit of the CNOT gate whose location is the even number in the linked CNOT chain
and values are reverse each other in the symmetic basis pair
for the controll qubit of the CNOT gate whose location is the odd number in the linked CNOT chain.
With this effect, the entanglement is dissolved in the eigenvalue output step 
for some code qubits in the symmetric basis pair.
This causes the decrease of the number of the gate in the entanglement dissolution step.

An example of the circuit for the 5 qubit code which is shown in Fig.\ref{fig:5qbleast}.
The number of the gates in this circuit is the smallest.
In Fig.\ref{fig:5qbleast}, first 4 gates form a linked CNOT chain.
\section{Summary}
This paper proposed more efficient decoding than the conventional decoding for some codes.
The proposed decoding outputs a part of the eigenvalues into a part of the code qubits
in contrast to the conventional method which output into the ancilla.
Besides, the proposed decoding dissolves a part of the entanglement in the eigenvalue output step
by using the linked CNOT gate in contrast to conventional method 
which does not dissolve in the eigenvalue output step.
As a result, the number of gates for the 5 qubit code was reduced to 6 
in contrast to 10 gates by the conventional method.
\bibliographystyle{junsrt}
\bibliography{myrefs,general,errorcorrection}
\end{document}